\documentclass[preprint,12pt]{elsarticle}

\usepackage{lineno,hyperref}
\usepackage{graphicx}
\usepackage{booktabs}
\usepackage{amssymb}
\usepackage{graphics}
\usepackage{amsmath}
\usepackage{placeins}
\usepackage{hyperref}
\usepackage{verbatim}
\usepackage[dvipsnames]{xcolor}
\usepackage{geometry}
\journal{Nuclear Instruments and Methods in Physics Research Section A}

\begin{document}

\begin{frontmatter}

\title{Development of a continuous-wave photocathode very-high-frequency electron gun for $\rm S^3FEL$}

\author[mymainaddress,mysecondaryaddress]{Lianmin Zheng}
\author[mymainaddress,mysecondaryaddress]{Yitong Duan}
\author[mymainaddress,mysecondaryaddress]{Yuanyuan Qin}
\author[mymainaddress,mysecondaryaddress]{Baiting Song}
\author[mymainaddress,mysecondaryaddress]{Yang Yu}
\author[mymainaddress,mysecondaryaddress]{Zixuan Dong}
\author[mymainaddress,mysecondaryaddress]{Yijiang Zhu}
\author[mymainaddress,mysecondaryaddress]{Yanqing Jia}
\author[mymainaddress,mysecondaryaddress]{Jiaru Shi}%
\author[mymainaddress,mysecondaryaddress]{Renkai Li}%
\author[mymainaddress,mysecondaryaddress]{Wenhui Huang}
\author[mymainaddress,mysecondaryaddress]{Huaibi Chen}%
\author[mymainaddress,mysecondaryaddress]{Chuanxiang Tang}%
\author[mymainaddress,mysecondaryaddress]{Yingchao Du\corref{mycorrespondingauthor}}
\ead{dych@tsinghua.edu.cn}
\cortext[mycorrespondingauthor]{Corresponding author}

\address[mymainaddress]{Department of Engineering Physics, Tsinghua University, Beijing 100084, People's Republic of China}
\address[mysecondaryaddress]{Key Laboratory of Particle and Radiation Imaging, Tsinghua University, Ministry of Education, Beijing 100084, People's Republic of China}

\begin{abstract}

A very-high-frequency (VHF) electron gun operating at a resonant frequency of 216.667 MHz has been developed at Tsinghua University for the Shenzhen Superconducting Soft X-ray Free-Electron Laser ($\rm S^3FEL$) facility. Building upon the SHINE gun design, the cavity profile was optimized to achieve a higher cathode electric-field gradient and a higher accelerating voltage at comparable input power. The revised cavity geometry also yields improved multipacting performance. In addition, thermal analysis was carried out to guide the water-cooling design, and enhanced cooling was implemented in the vicinity of the cathode. During high-power conditioning, 82 kW of continuous-wave radio-frequency power was successfully coupled into the gun, corresponding to a cathode gradient of 29.9 MV/m and an accelerating voltage of 856 kV. The gun voltage surpasses the previous world record for room-temperature VHF guns. The maximum dark current measured by the Faraday cup at the gun exit was only 6.8 nA.
\end{abstract}


\end{frontmatter}


\section{\label{sec:level1}Introduction}

A very-high-frequency (VHF) photocathode electron gun~\cite{Sannibale2012APEX, wells2016mechanical, sannibale2017upgrade, Sannibale2019VHFgun, luo2019rf, zheng2023design, zheng2019design, chen2022beam, chen2021analysis, wang2025experimental, Gu2020_162.5MHzGun} is an electron source capable of producing high-quality electron beams at megahertz repetition rates.
It serves as an ideal electron source for high-repetition-rate free-electron lasers (FELs) and has been adopted as the electron source at facilities such as LCLS-II~\cite{stohr2011linac, zhou2023overview} in the United States and SHINE~\cite{yan2019overview, zhao2024shanghai} in China.
In addition, VHF electron guns have found important applications in high-repetition-rate ultrafast electron diffraction (UED)~\cite{filippetto2016design, tan2026z, song2026time} and ERL-FEL-based lithography light sources~\cite{kawata2022high, chen2025design}.

The first VHF electron gun developed at Tsinghua University~\cite{zheng2023design} has been successfully deployed at SHINE and the Dalian Advanced Light Source Test Facility~\cite{sun2022conceptual, li2025injector}. The gun operates at a resonant frequency of 216.667 MHz in continuous-wave (CW) radio-frequency (RF) mode. During high-power conditioning, it achieved a cathode gradient of 27 MV/m and a voltage of 780 kV at an input power of 75 kW. It holds the record for the highest cathode gradient and voltage among room-temperature CW electron guns.

Shenzhen Superconducting Soft X-ray Free-Electron Laser ($\rm S^3FEL$)~\cite{wang2023physical, Zhang2023_S3FELphotoinjector} is China's second high-repetition-rate free-electron laser facility approved for construction, following the SHINE facility in Shanghai. It is designed for a beam energy of 2.5 GeV and a beam repetition rate of 1 MHz. Like SHINE, it employs a VHF electron gun developed by Tsinghua University as its electron source. This provides a valuable opportunity to upgrade our electron gun design. The electron gun has been developed and successfully passed high-power testing. During testing, it achieved a cathode electric field gradient of 29.9 MV/m and a voltage of 856 kV at an input power of 82 kW. The gun voltage once again broke the previous world record held by our team. This paper describes the design improvements of the new gun compared to the SHINE/DALS gun, as well as the results of its cold tests and high-power tests.

\section{RF and multipacting optimization}

The resonant frequency of the new gun is identical to that of the SHINE/DALS gun, at 216.667 MHz. First, we optimized the cavity profile to achieve improved RF performance relative to the SHINE gun. A key objective of this optimization was to further increase the voltage while keeping the input power and cathode electric field gradient essentially unchanged. In FEL applications, higher voltages can suppress the adverse effects of space charge on beam quality during beam transport; when the new gun is applied to UED systems in the future, it can also enhance sample penetration and reduce multiple scattering.

\begin{figure}[htbp]
\begin{center}
\includegraphics[width=6cm]{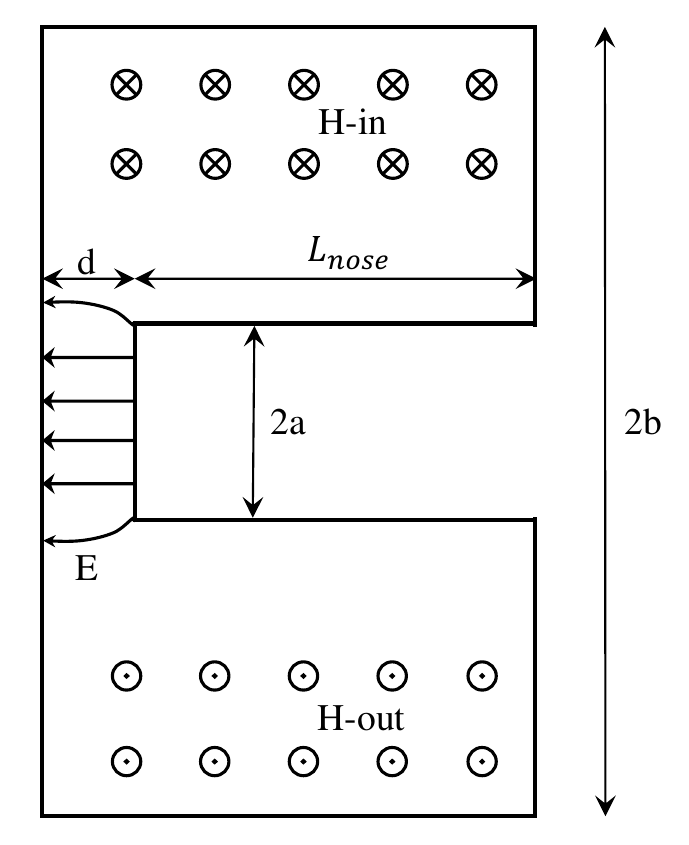}
\end{center}
\caption{ Simplified structural model of the electron gun cavity.}
\label{fig:1}
\end{figure}

If the cavity is modeled as an equivalent parallel RLC circuit, the resonant angular frequency can be expressed as $\omega_0 = 1/\sqrt{LC}$. The characteristic impedance $R_{\text{sh}}/Q_0 = \omega_0 L$ is a quantity determined solely by the cavity geometry, where $R_{\text{sh}}$ is the shunt impedance and $Q_0$ is the unloaded quality factor. To increase the voltage at the same input power, $R_{\text{sh}}$ must be increased, which requires increasing $L$. At the same time, to keep the resonant frequency constant, $C$ must be reduced accordingly. To establish a relationship between $L$, $C$, and the cavity profile dimensions, we simplified the cavity structure to the form shown in Fig.~\ref{fig:1}. In this simplified model, the capacitance and inductance are separated. The region where the electric field is strongest can be modeled as a parallel-plate capacitor with $C = \epsilon_0 \pi a^2 / d$. The remaining portion can be modeled as a coaxial inductor with $L = (\mu_0 / 2\pi) \, L_{\text{nose}} \, \ln(b/a)$. It can be seen that to increase $L$ while simultaneously decreasing $C$, one can increase the anode–cathode gap $d$, increase the cavity radius $b$, reduce the cathode nose cone radius $a$, or increase the cathode nose cone length $L_{\text{nose}}$.

\begin{figure}[htbp]
\begin{center}
\includegraphics[width=8.5cm]{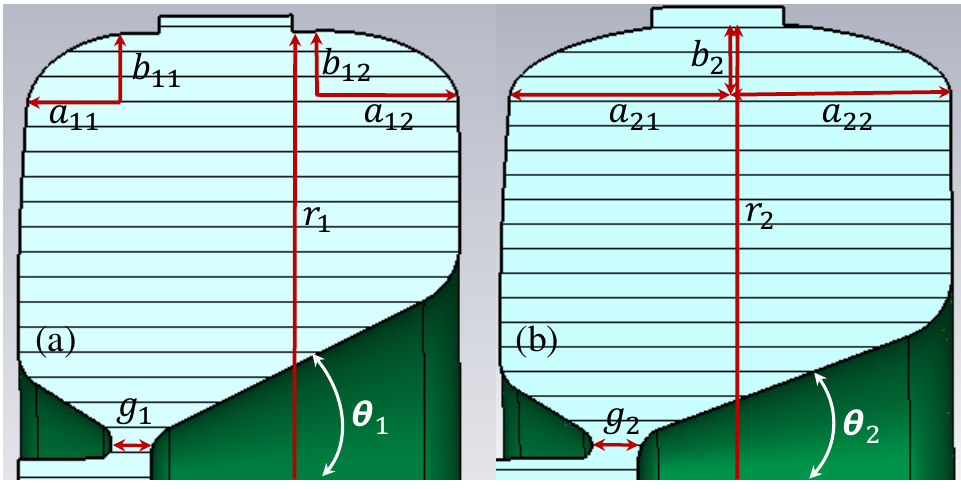}
\end{center}
\caption{(a) The cavity profile of the SHINE gun, (b) the optimized cavity profile of the new gun.}
\label{fig:2}
\end{figure}

Based on the above model, the direction for cavity optimization was determined. The optimized cavity profile is shown in Fig.~\ref{fig:2}; the cavity profile of the SHINE gun is also shown for comparison. To achieve a higher shunt impedance $R_{\text{sh}}$, dimensional changes were made primarily in the following three areas: (1) the anode–cathode gap was increased from 3~cm to 3.5~cm; (2) the cavity radius was increased from 33.5~cm to 34.7~cm; (3) the cathode nose cone angle was reduced from 26$^\circ$ to 20$^\circ$. The remaining dimensions were parameterized and optimized using CST Microwave Studio~\cite{cst}. The optimization objectives included the gun voltage, input power, cathode gradient, peak surface power density, and peak surface electric field.

After multiple rounds of combined optimization of RF performance and multipacting behavior, the final key dimensions of the cavity profile were determined and are presented in Table~\ref{tab:table1}.

\begin{table}[]
\caption{\label{tab:table1} Key dimensions of the finalized cavity profile.}
\begin{tabular}{@{}lcc@{}}
\toprule
& SHINE/DALS gun& new gun         \\ \midrule
Anode-cathode gap &   $g_1=30$ mm  &   $g_2=35$ mm           \\
Cathode nose cone inclination angle &    $\theta_1=26^{\circ}$ &   $\theta_2=20^{\circ}$   \\
Cavity radius &  $r_1=33.5$ cm & $r_2=34.7$ cm    \\
\raisebox{-\height}{Dimensions of outer cavity profile} & \begin{tabular}[t]{@{}c@{}}$a_{11}=$8.38 mm\\ $a_{12}=$10 mm\\$b_{11}=$6 mm\\ $b_{12}=$5.7 mm\end{tabular}            &  \begin{tabular}[t]{@{}c@{}}$a_{21}=$16.3 cm\\ $a_{22}=$17.2 cm\\ $b_{2}=$5.7 cm\end{tabular}
\\ \bottomrule
\end{tabular}
\end{table}

The optimized RF and multipacting performance are as follows.

\subsection{Optimized RF performance}

During the RF parameter optimization, we initially assumed that the top surface of the cathode plug is flush with that of the cathode nose. Under this condition, the design values of the optimized RF parameters for the new gun are listed in Table~\ref{tab:table2}. For comparison, the measured RF parameters obtained from the high-power test of the SHINE/DALS gun are also included in the table. As shown, at the same cathode gradient of 27 MV/m, the gun voltage was increased from 780 kV to 852 kV.

\begin{figure}[htbp]
\begin{center}
\includegraphics[width=8cm]{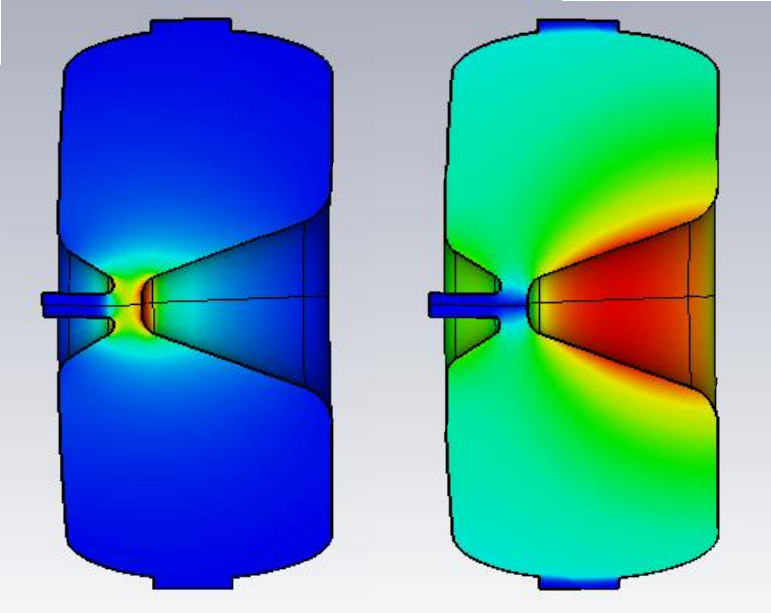}
\end{center}
\caption{ Electric (left) and magnetic (right) field distributions of the new gun.}
\label{fig:3}
\end{figure}

The simulation shows that achieving a voltage of 852 kV requires 76.2 kW of input power. However, because the quality factor of the assembled gun is typically slightly lower than the design value, a somewhat higher input power is required in practice. The electric and magnetic field distributions of the new gun are depicted in Fig.~\ref{fig:3}.

\begin{table}[]
\caption{\label{tab:table2} Design values of the optimized RF parameters for the $\rm S^3$FEL gun. The corresponding values measured in the high-power test of the SHINE/DALS gun are also listed for comparison. Two operating modes are considered: in the ``normal plug'' mode, the top surface of the cathode plug is flush with that of the cathode nose; in the other mode, the plug is inserted by an additional 0.5 mm.}
\begin{tabular}{@{}lcccr@{}}
\toprule
RF Parameters               & \begin{tabular}[c]{@{}c@{}}Achieved value for \\ SHINE/DALS gun\\(normal plug)\end{tabular} & \begin{tabular}[c]{@{}c@{}}Design value for \\ $\rm S^3$FEL gun\\(normal plug)\end{tabular} & \begin{tabular}[c]{@{}c@{}}Design value for \\ $\rm S^3$FEL gun\\(0.5-mm plug \\over-insertion)\end{tabular}& Unit         \\ \midrule
Frequency                   &   216.667  &   216.667    & 216.667  & MHz  \\
Cathode gradient            &    27     &   27 &  29.86  & MV/m         \\
Input power                 &  75 & 76.2 & 76.2  & kW           \\ 
Peak surface power density  &  23.0    &   29.5    &  29.5  & $\rm W/cm^2$ \\
Voltage                     & 780 & 852 & 854 & kV           \\
Quality factor              &  33717& 34545     &34533  &              \\
Shunt impedance             &  8.34& 9.53 &9.56 & M$\Omega$    \\ \bottomrule
\end{tabular}
\end{table}


\begin{figure}[htbp]
\begin{center}
\includegraphics[width=7.5cm]{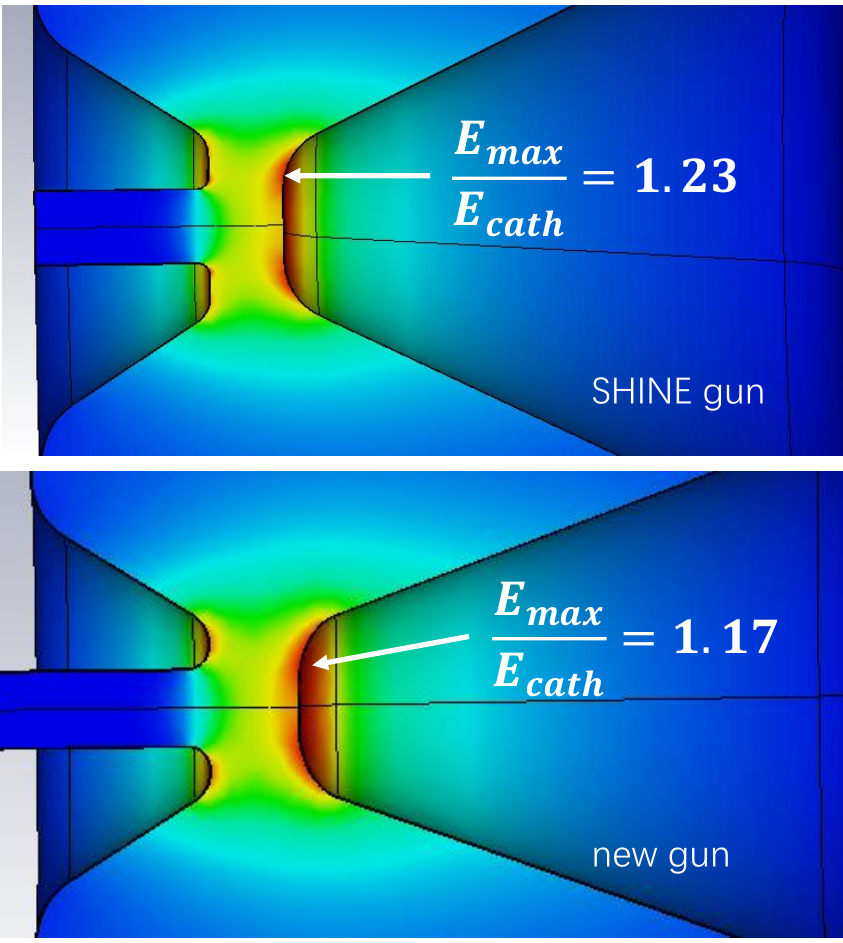}
\end{center}
\caption{ Electric field distribution at the anode-cathode gap: Top image shows the SHINE gun; Bottom image shows the new gun.}
\label{fig:4}
\end{figure}

The arc at the top of the cathode nose corresponds to the region of highest surface electric field. To reduce the maximum surface electric field and thereby lower the breakdown probability, we optimized the contour of this region, as shown in Fig.~\ref{fig:4}. For the SHINE gun, the radius of the circular arc was 25 mm. We replaced this circular arc with an elliptical arc and optimized its major and minor axes. The optimized elliptical arc had a major axis of 28.3 mm and a minor axis of 19.4 mm. The optimized local maximum electric field was significantly reduced. $E_{\max}/E_{\text{cath}}$ decreased from 1.23 for the SHINE gun to 1.17.

\begin{figure}[htbp]
\begin{center}
\includegraphics[width=9cm]{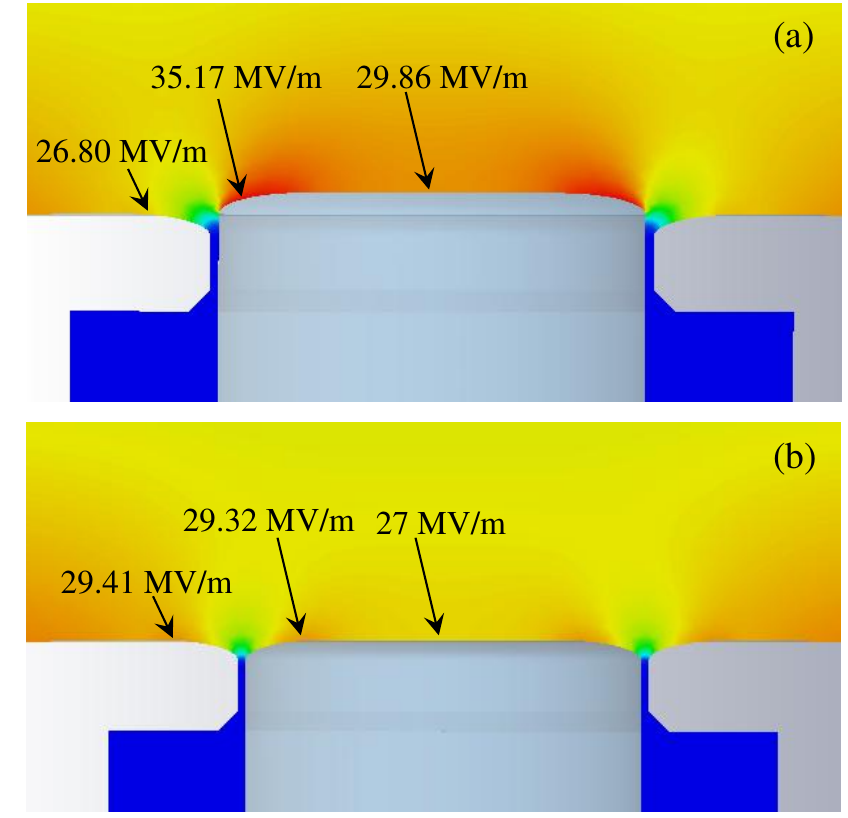}
\end{center}
\caption{Electric field distributions at the corners of the joints of the cathode plug and the cathode nose base plate with (a) 0.5 mm over-insertion of the plug and (b) normal insertion of the plug. The input rf power is 76.2 kW.}
\label{fig:plugoverinsert}
\end{figure}

In the above simulation, the joint structures of the cathode plug and the cathode nose base plate were not considered. Reference~\cite{wang2025experimental} proposed a novel cathode plug over-insertion scheme to suppress dark current, and the effectiveness of this method was experimentally verified. Subsequently, this method was applied to the SHINE gun and the DALS gun, reducing the dark current to the order of several nA. During the design phase of the $\rm S^3$FEL gun, we also considered this over-insertion scheme and set the over-insertion depth to 0.5 mm. The corners of the joints between the cathode plug and the copper base plate of the $\rm S^3$FEL gun feature an elliptical arc design, with a major axis of 2 mm and a minor axis of 0.5 mm. Figure~\ref{fig:plugoverinsert} illustrates the electric field distribution near the cathode plug under two operating modes: 0.5 mm over-insertion and normal insertion. It can be seen that in the 0.5 mm over-insertion scheme, the electric field at the plug corner is significantly enhanced, while the field at the cathode nose base plate is substantially weakened. The electric field diverges along the beam direction, causing the dark current to spread out rapidly and be lost in the beam pipe. Compared to normal insertion, a 0.5 mm over-insertion results in a decrease in the resonant frequency of approximately 8 kHz. We fine-tuned the anode–cathode gap to maintain the gun's resonant frequency at 216.667 MHz. Under these conditions, the RF parameters for the same input power of 76.2 kW are listed in Table~\ref{tab:table2}. It can be seen that the cathode electric field gradient increased significantly, from 27 MV/m to 29.86 MV/m, while the other RF parameters remained essentially unchanged.

\subsection{Optimized multipacting performance}

In the SHINE/DALS gun, simulations and high-power tests have shown that multipacting occurs when the cathode gradient is below 16~MV/m, but ceases once the gradient exceeds this threshold. Although multipacting does not occur at the operating cathode gradient, pulse mode must be used during the startup phase to bypass the multipacting regime. In pulse mode, both the rising and falling edges still traverse the multipacting region, causing vacuum degradation inside the gun and compromising the service lifetime of the semiconductor cathode. Therefore, there remains a strong incentive to further suppress the intensity of multipacting.

Building on the SHINE/DALS gun, we further optimized the multipacting performance. In Ref.~\cite{zheng2023design}, we demonstrated that reducing the cavity radius and modifying the outer contour from circular to elliptical can significantly reduce the multipacting intensity. However, the new gun features a larger cavity radius to increase the shunt impedance; consequently, if the original outer contour were retained, multipacting would inevitably become more severe. To mitigate multipacting, we proposed an equatorial cavity configuration that maximizes the major axis of the ellipses and eliminates the straight segment of the outer contour (Fig.~\ref{fig:2}(b)). We thoroughly investigated the physical mechanism by which this structure suppresses multipacting and reported our findings in Ref.~\cite{duan2026toward}.

\begin{figure}[htbp]
\begin{center}
\includegraphics[width=9cm]{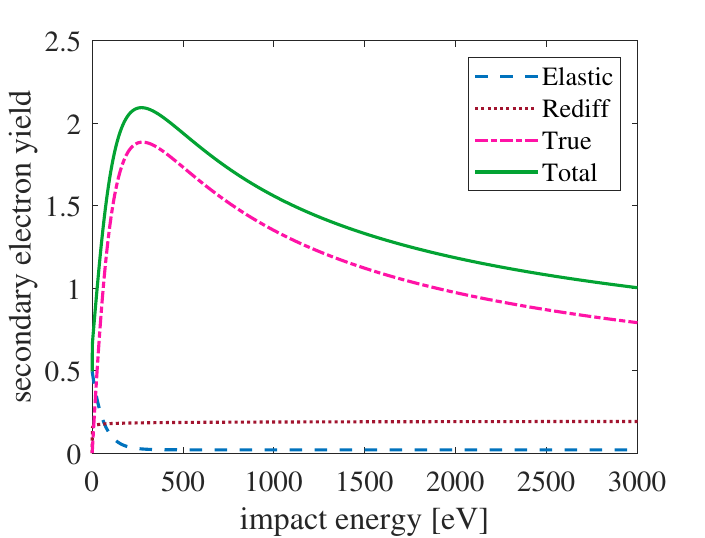}
\end{center}
\caption{Secondary electron yield of copper as a function of impact energy.}
\label{fig:sey}
\end{figure}

The multipacting process was simulated in CST Particle Studio using a tracking solver. The RF field distribution was imported from CST Microwave Studio. The secondary electron yield (SEY) was calculated using the built-in Furman model in CST. The Furman model is a sophisticated physical framework that accurately describes electron–surface interactions. It divides secondary electron emission into three components: true secondary electrons, rediffused electrons, and elastically backscattered electrons. The total secondary electron emission is the sum of these three components. The SEY of copper as a function of impact energy is shown in Fig.~\ref{fig:sey}.

\begin{figure}[htbp]
\begin{center}
\includegraphics[width=9cm]{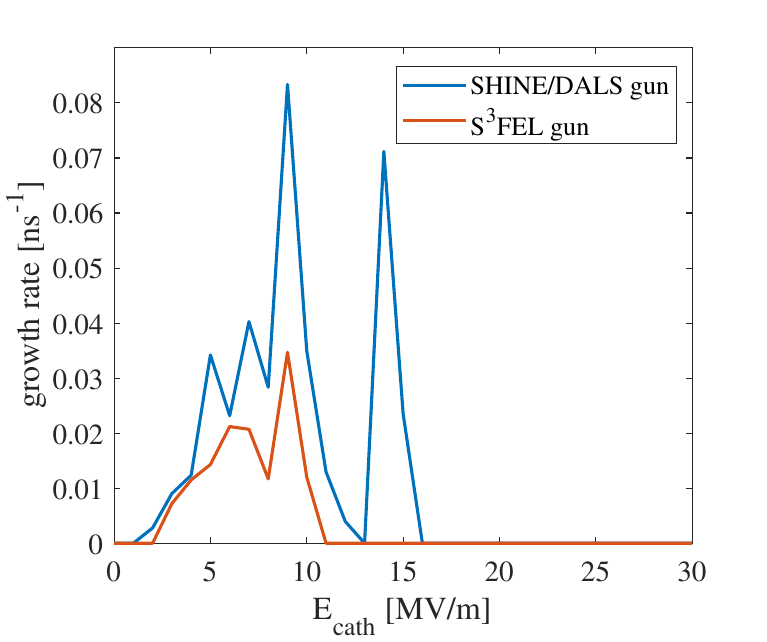}
\end{center}
\caption{ growth rate $\alpha$ of the $\rm S^3$FEL gun and the SHINE/DALS gun as a function of the cathode gradient $E_{cath}$.}
\label{fig:growth_rate}
\end{figure}

Initially, seed electrons are generated on the inner wall of the cavity, and the evolution of the total electron population over time is tracked under various cathode gradient conditions. If the number of particles grows exponentially over time—indicating that multipacting has occurred in the gun—the growth curve can be fitted with the equation $N_e(t) = N_0 e^{\alpha t}$, where $N_e(t)$ is the particle number at time $t$, $N_0$ is the initial particle number, and $\alpha$ is the growth rate, which characterizes the multipacting intensity. Conversely, if the particle count decreases over time, it indicates that no multipacting has occurred. The growth rate $\alpha$ of the $\rm S^3$FEL gun as a function of the cathode gradient is shown in Fig.~\ref{fig:growth_rate}. For comparison, the multipacting intensity of the SHINE/DALS gun is also plotted. It is evident that both the multipacting intensity and the multipacting range of the new gun have been significantly reduced. No multipacting is observed when the cathode gradient exceeds 11~MV/m. This result is highly favorable for the stable operation of the electron gun.

\section{Mechanical structure and thermal analysis}

\begin{figure}[htbp]
\begin{center}
\includegraphics[width=9cm]{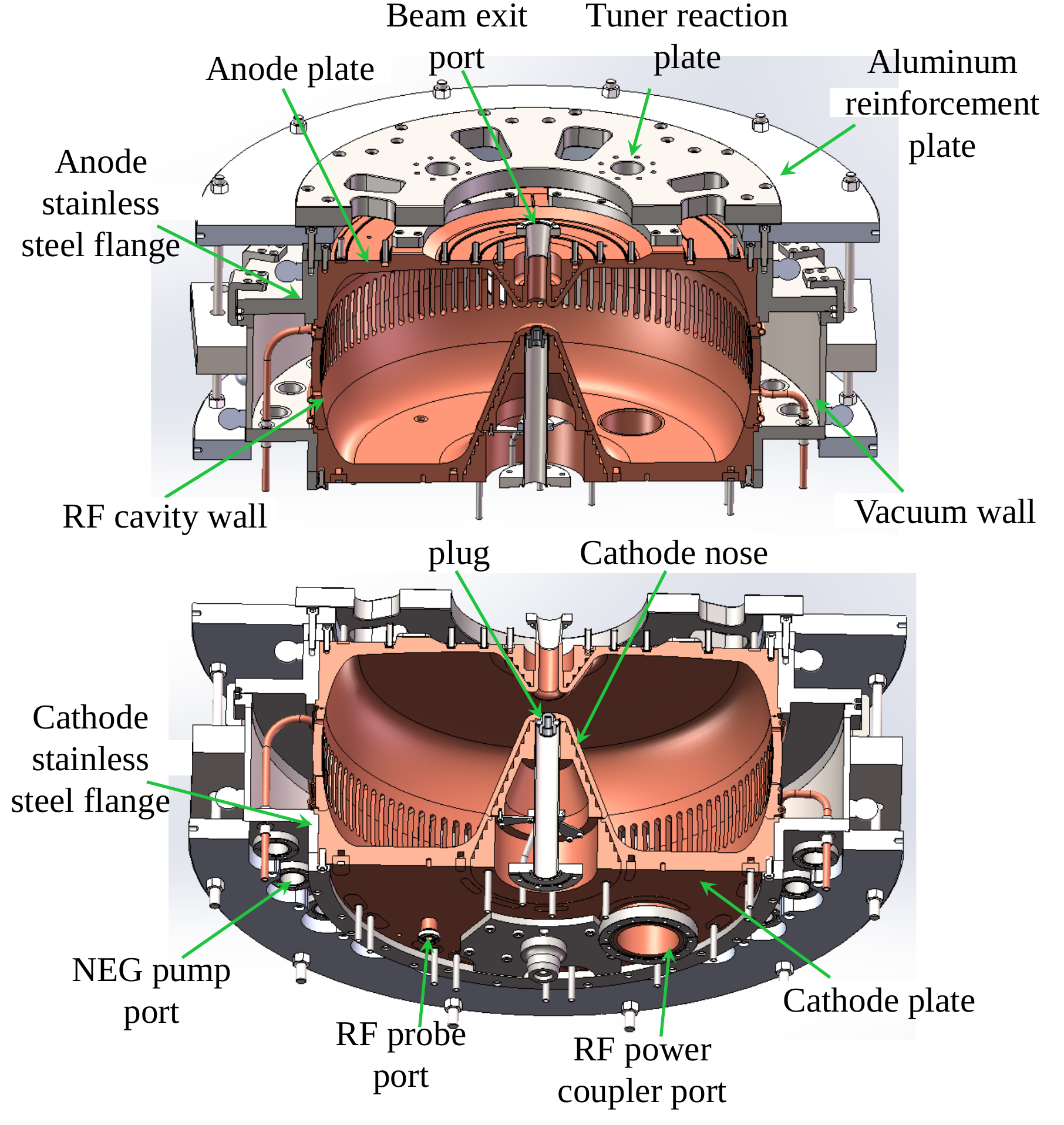}
\end{center}
\caption{Mechanical structure of the $\rm S^3$FEL gun.}
\label{fig:gun_image}
\end{figure}

The mechanical structure of the gun is illustrated in Fig.~\ref{fig:gun_image}. This structure is similar to that of the SHINE/DALS gun, except that the copper cavity dimensions were modified to accommodate the changes in the RF cavity shape. The gun is fabricated from oxygen-free copper and 316L non-magnetic stainless steel. The fabrication process is essentially identical to that of the SHINE/DALS gun.

\begin{figure}[htbp]
\begin{center}
\includegraphics[width=9cm]{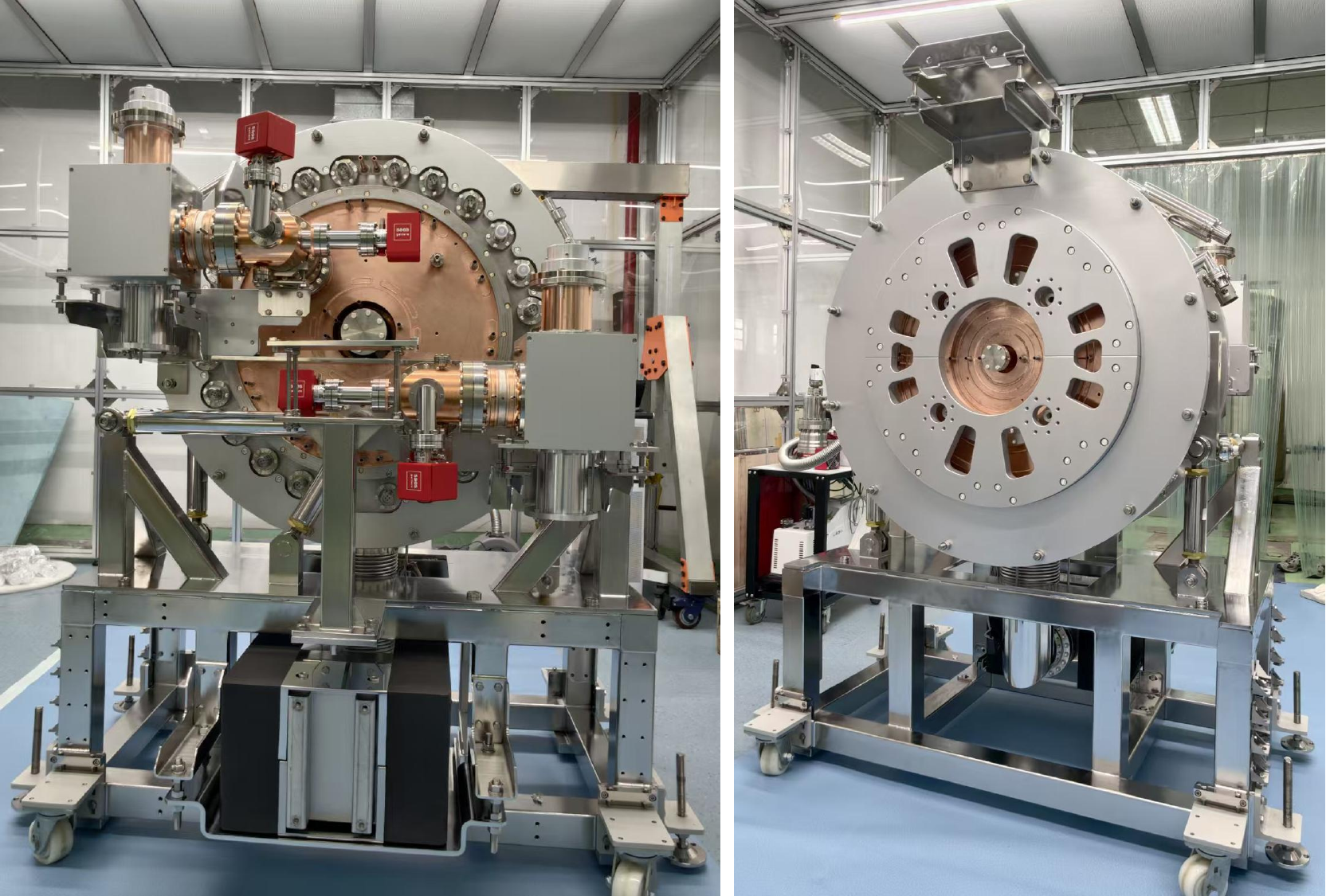}
\end{center}
\caption{Photographs of the electron gun after assembly. Left: cathode side. Right: anode side.}
\label{fig:gun_image_assemble}
\end{figure}

Figure~\ref{fig:gun_image_assemble} shows photographs of the assembled $\rm S^3$FEL gun. On the cathode stainless steel flange, there are 24 CF~35 flanges, onto which 24 SAES CapaciTorr Z400 NEG pumps were installed. On the vacuum wall, there are six CF~35 flanges, accommodating a VAT rough-pumping angle valve, a MicroVision residual gas analyzer, and a Leybold IE514 vacuum gauge. Below the vacuum wall, there is a CF~150 flange port connected to a 500~L/s Agilent ion pump. Above the vacuum wall, there is a CF~100 flange port connected to a 100~L/s Agilent ion pump (not shown in Fig.~\ref{fig:gun_image_assemble}, but visible in Fig.~\ref{fig:high_power_test_image}). In addition, two SAES NexTorr NEG-ion Z200 pumps are installed on the waveguide of each coupler.

\begin{figure}[htbp]
\begin{center}
\includegraphics[width=9cm]{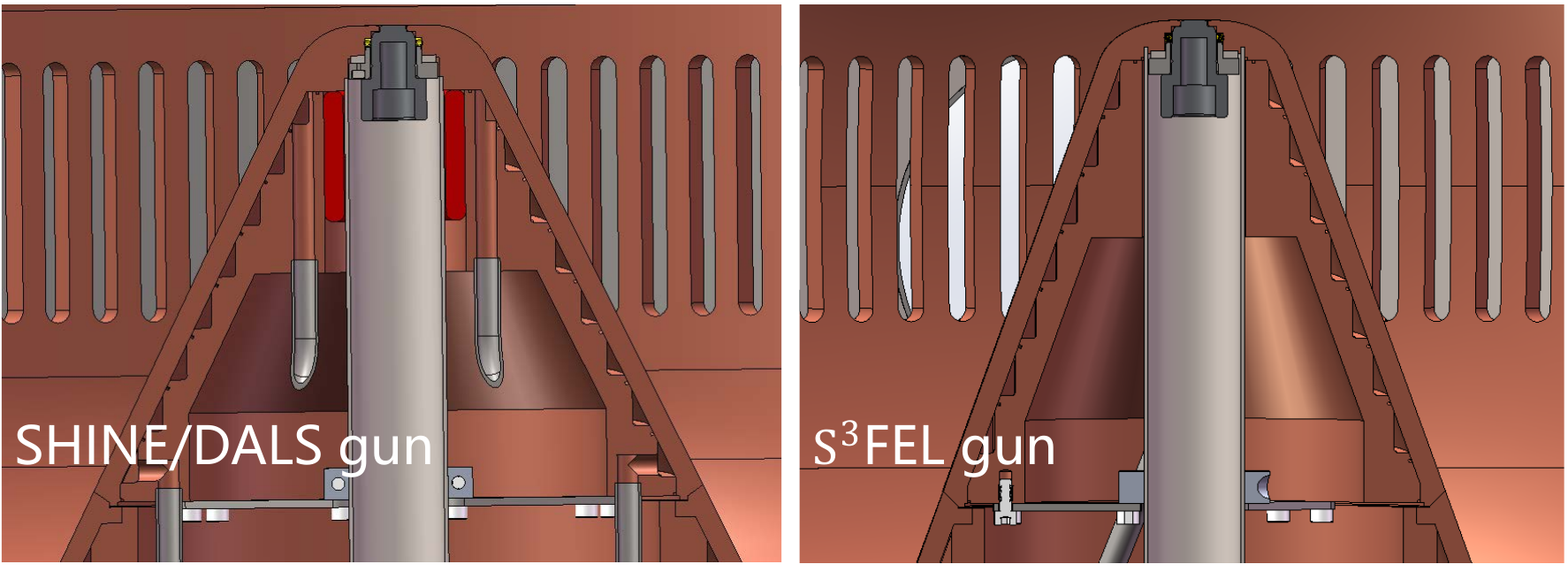}
\end{center}
\caption{Cathode nose structure of the SHINE/DALS gun (left) and the $\rm S^3$FEL gun (right).}
\label{fig:nose_structure}
\end{figure}

A bucking solenoid was installed in the cathode nose of the SHINE/DALS gun to cancel any residual magnetic field from the main solenoid at the cathode. However, in both the SHINE and DALS injectors, the main solenoids were designed to leave no residual magnetic field at the cathode; therefore, this bucking solenoid was not activated. After consulting with the $\rm S^3$FEL project team, we decided to remove this bucking solenoid. The cathode nose structures of the SHINE/DALS gun and the $\rm S^3$FEL gun are shown in Fig.~\ref{fig:nose_structure}. One reason for this decision is that the cathode nose of the $\rm S^3$FEL gun is narrower, making it difficult to accommodate the bucking solenoid. Additionally, removing the bucking solenoid allows the two spiral water channels inside the cathode nose to be positioned closer to the nose tip.

\begin{figure}[htbp]
\begin{center}
\includegraphics[width=9cm]{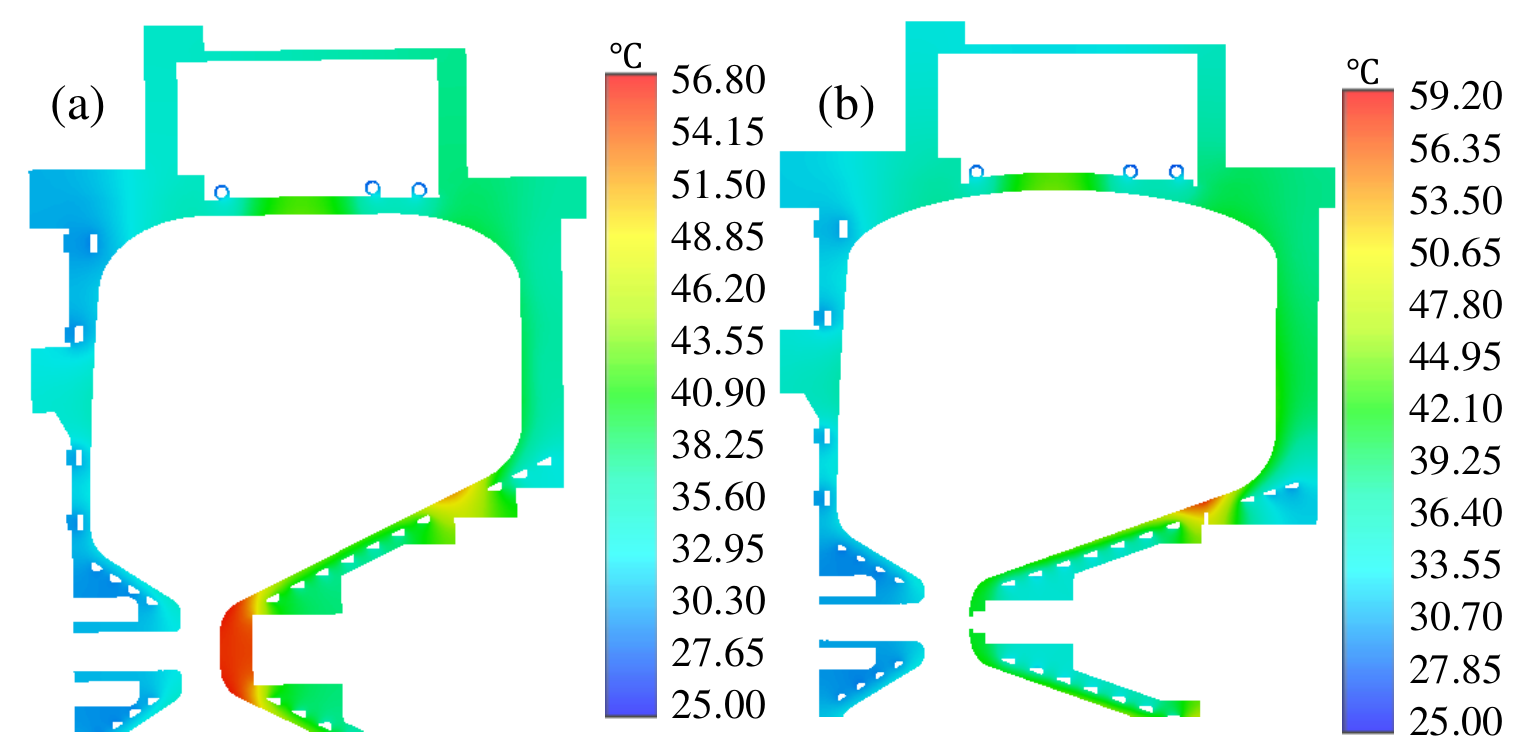}
\end{center}
\caption{The steady-state temperature distribution of the SHINE/DALS gun (left) and the $\rm S^3$FEL gun (right). Both types of guns have an input power of 76 kW.}
\label{fig:gun_temp}
\end{figure}

We built an electromagnetic–thermal–structural finite element analysis model in ANSYS to study the effects of RF heating on the electron gun. The gun contains a total of 23 water channels: 6 in the anode plate, 12 in the cathode plate, 2 in the cathode nose, and 3 on the RF cavity wall. The flow rate was set to 15~L/min for the two cooling water channels in the cathode nose and to 10~L/min for the remaining channels. The initial cooling water temperature was set to 25~$^\circ$C. The input power was set to 76~kW. The steady-state temperature distribution obtained from the simulation is shown in Fig.~\ref{fig:gun_temp}. For comparison, the temperature distribution of the SHINE/DALS gun under the same input power and water flow conditions is also shown. Compared to the SHINE/DALS gun, the temperature at the cathode nose tip of the new gun dropped significantly, from 57~$^\circ$C to 42~$^\circ$C, which helps extend the lifetime of the semiconductor cathode. The highest temperature in the new gun occurs at the junction between the cathode nose and the cathode plate, reaching 59.2~$^\circ$C. The power density on the cathode nose of the new gun is higher than that of the SHINE/DALS gun; therefore, the temperature at this location is higher in the new gun than in the SHINE/DALS gun.

\begin{figure}[htbp]
\begin{center}
\includegraphics[width=9cm]{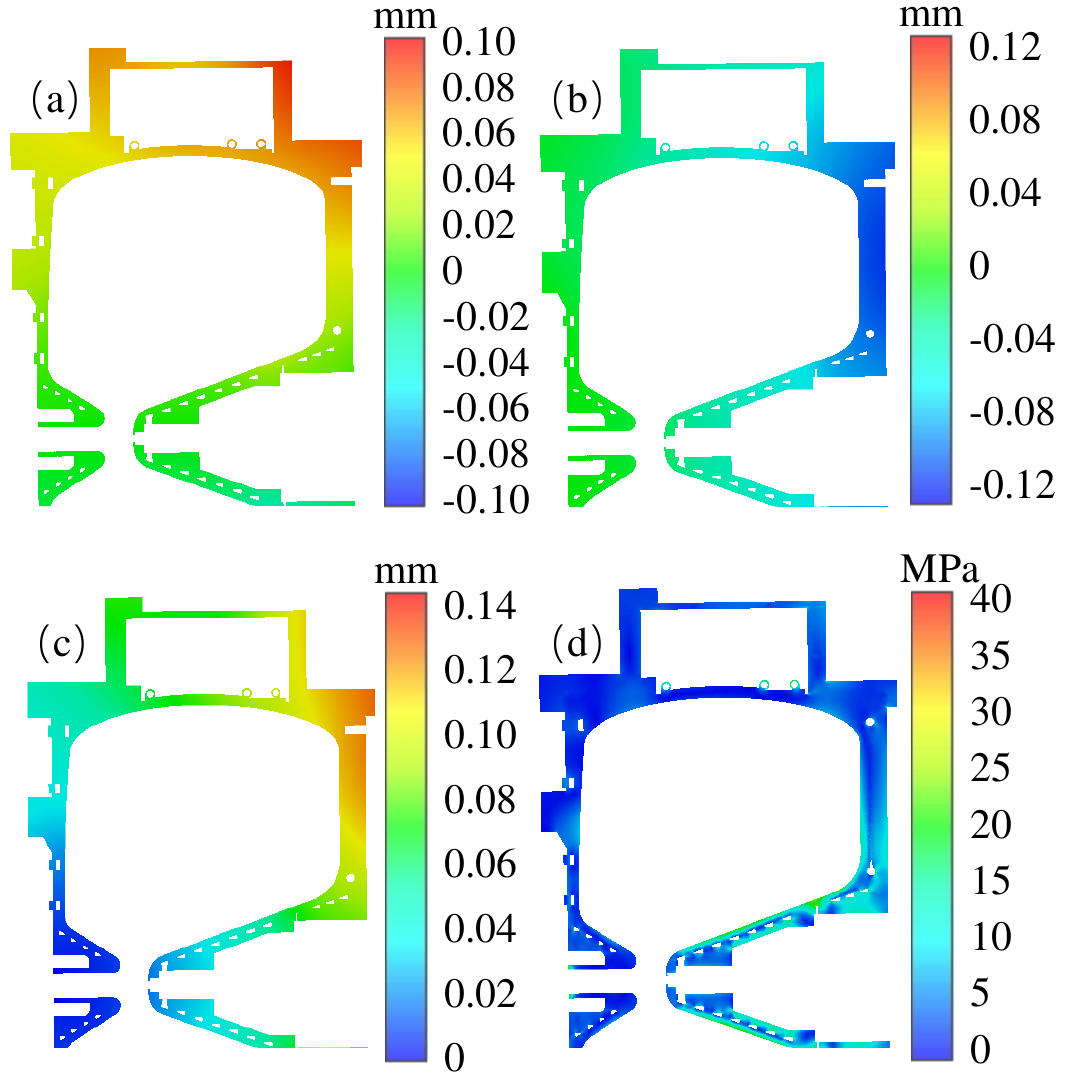}
\end{center}
\caption{Radial (a), axial (b) and total (c) deformation of the $\rm S^3$FEL gun due to thermal expansion. The distribution of the von Mises equivalent stress is shown in (d).}
\label{fig:displacement_total}
\end{figure}

Based on this, we simulated the deformation of the electron gun caused by thermal expansion, as shown in Fig.~\ref{fig:displacement_total}. The end face of the beam exit port at the center of the anode plate was treated as a fixed surface, and the electron gun was allowed to expand freely. The gun undergoes radial thermal expansion, with a maximum displacement of 100~$\mu$m, leading to a decrease in frequency. In the axial direction, thermal expansion of the RF cavity wall causes the cathode plate to retreat by 120~$\mu$m relative to the anode plate, while the cathode nose moves forward relative to the cathode plate. The net effect is an increase in the anode–cathode gap of approximately 17~$\mu$m. The stress distribution resulting from the thermal effect is shown in Fig.~\ref{fig:displacement_total}(d). The stresses are below 40~MPa, which is lower than the yield strength of copper. Simulations indicate that the frequency shift due to thermal effects under 76~kW of RF power is $-$46~kHz.

\section{Cold test}

\begin{figure}[htbp]
\begin{center}
\includegraphics[width=8.5cm]{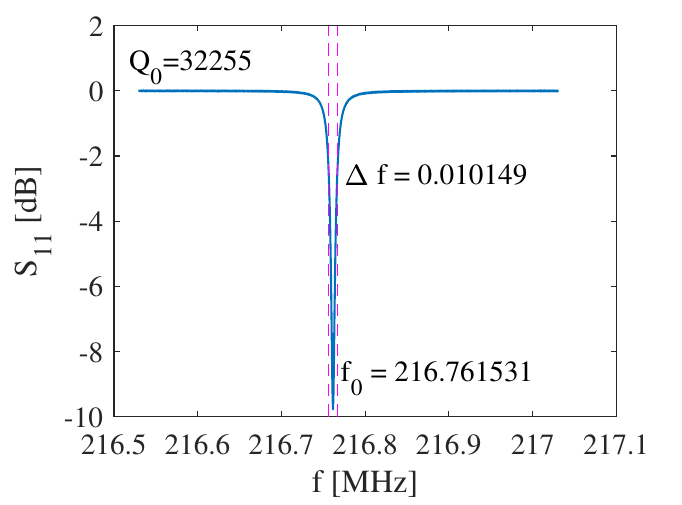}
\end{center}
\caption{cold test of the electron gun with one coupler port connected to the vector network analyzer and the other coupler port short-circuited.}
\label{fig:S11}
\end{figure}

After assembling the electron gun, we performed a cold test of the resonant frequency and quality factor using a vector network analyzer, as shown in Fig.~\ref{fig:S11}. During the measurement, one coupler port was connected to the vector network analyzer, while the other coupler port was terminated with a short. The measurement temperature was 24~$^\circ$C, and the interior of the gun was at atmospheric pressure. The unloaded quality factor was measured to be 32,255, corresponding to 93.3\% of the design value. The coupling factors of the two couplers were measured to be 0.53 and 0.54, respectively. The coupling factors of the two RF probes were $-$62.89~dB and $-$61.19~dB, respectively.

\section{High power conditioning}

\begin{figure}[htbp]
\begin{center}
\includegraphics[width=8cm]{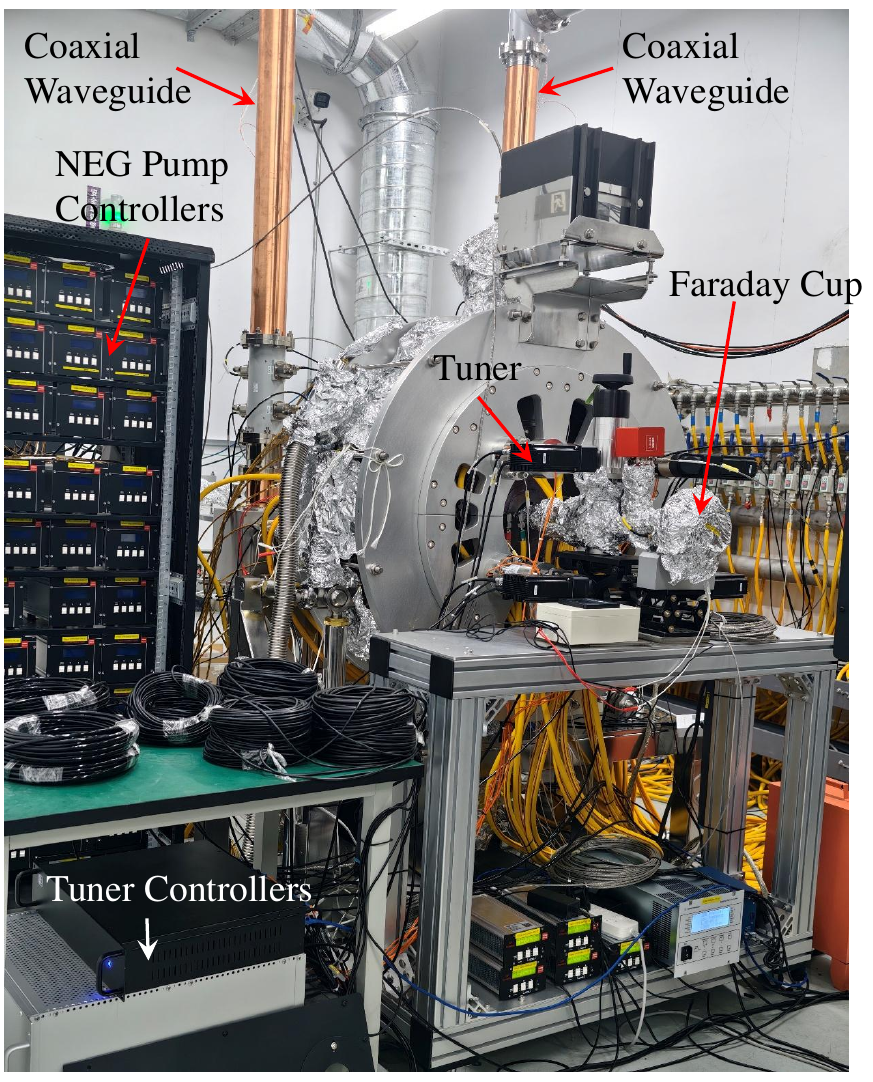}
\end{center}
\caption{On-site photo of high-power conditioning of the $\rm S^3$FEL gun.}
\label{fig:high_power_test_image}
\end{figure}

After completing the cold test, the gun was transported to a bunker at Tsinghua University's Changping Research Base for high-power conditioning. Fig.~\ref{fig:high_power_test_image} shows a photograph of the site after the waveguide, water circuit, tuners, and other instruments had been assembled. A simple loadlock was installed on the gun to insert a 316L stainless steel plug. The top surface of the plug is 0.5~mm above the top face of the cathode nose, serving to suppress dark current.

The power supply system of the electron gun consists of two 60-kW solid-state amplifiers (SSAs). Each SSA is connected to a coupler on the gun via a coaxial waveguide. During the initial power ramp-up, the microwave power was gradually increased to 68~kW in short-pulse mode. During this period, microcurrent detectors on the coupler waveguide detected current signals on the order of hundreds of nA, indicating that multipacting had occurred within the coupler waveguide. After several hours of conditioning, the microcurrent dropped to zero. Throughout this process, the water cooling in the electron gun was turned off, while the water cooling on the coupler remained on. The gun temperature gradually rose, and by adjusting the microwave duty cycle, the temperature was stabilized at approximately 120~$^\circ$C. After 80 hours of RF baking, we activated the NEG pumps, turned on the ion pumps, and applied water cooling to the gun. Without RF power, the vacuum inside the gun can be maintained at $4\times10^{-9}$~Pa.

\begin{figure}[htbp]
\begin{center}
\includegraphics[width=9cm]{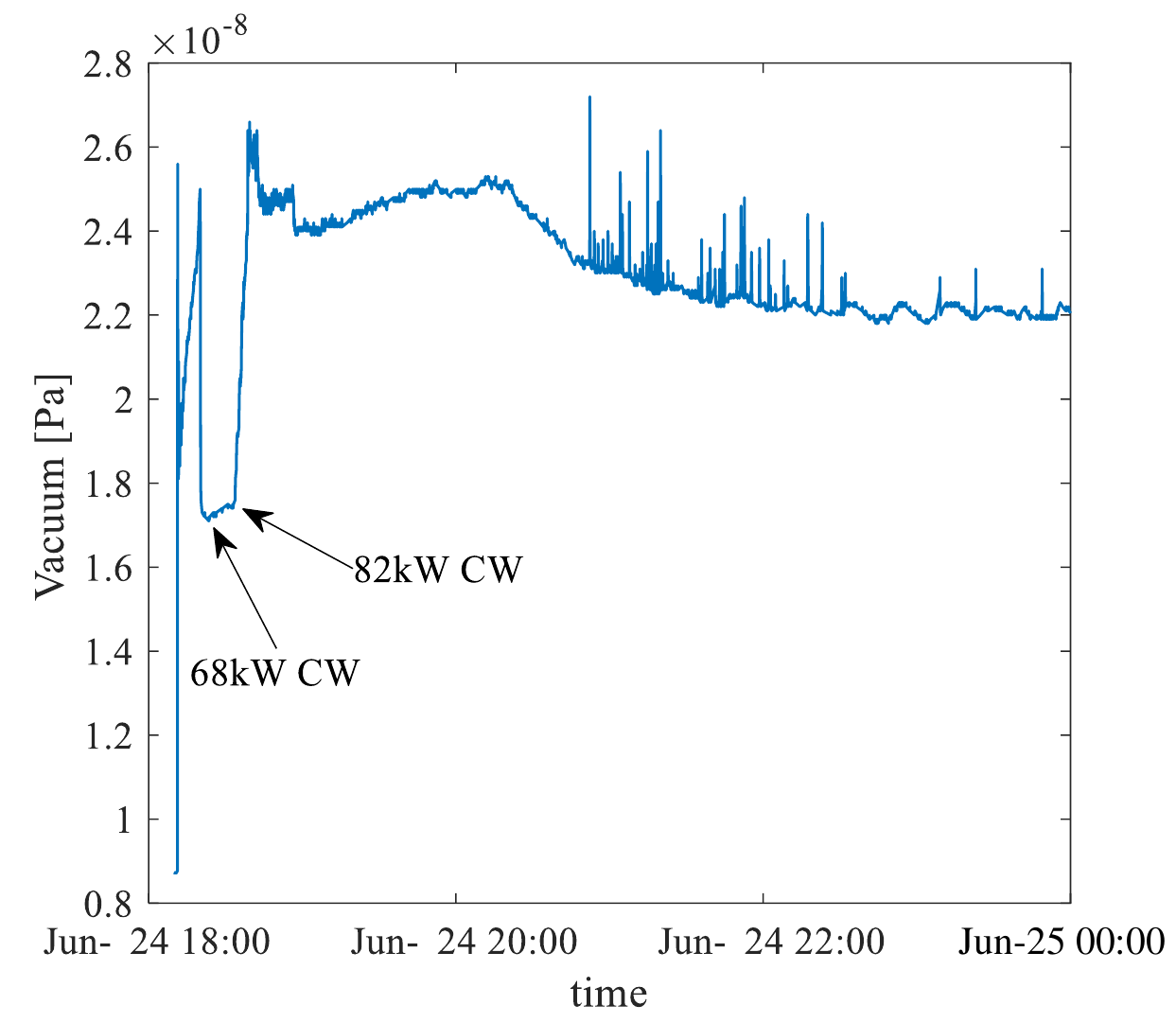}
\end{center}
\caption{Changes in vacuum during the high-power conditioning. A pulsed wave with a peak power of 68 kW gradually increased its duty cycle to CW, after which the power was increased to 82 kW and maintained for 5 hours.}
\label{fig:vacuum}
\end{figure}

Fig.~\ref{fig:vacuum} shows the vacuum variation during the first injection of 82~kW of CW RF power. Initially, the gun was operated in pulsed mode at 68~kW, and the pulse width was gradually increased. Upon switching to CW mode, a noticeable vacuum drop occurred. This behavior arises because the rising and falling edges of the pulses induced multipacting inside the gun, which ceased once the system entered CW operation. Subsequently, the CW power was increased to 82~kW and maintained for 5 hours. The vacuum in the gun eventually stabilized at $2.2\times10^{-8}$~Pa.

\begin{figure}[htbp]
\begin{center}
\includegraphics[width=8.5cm]{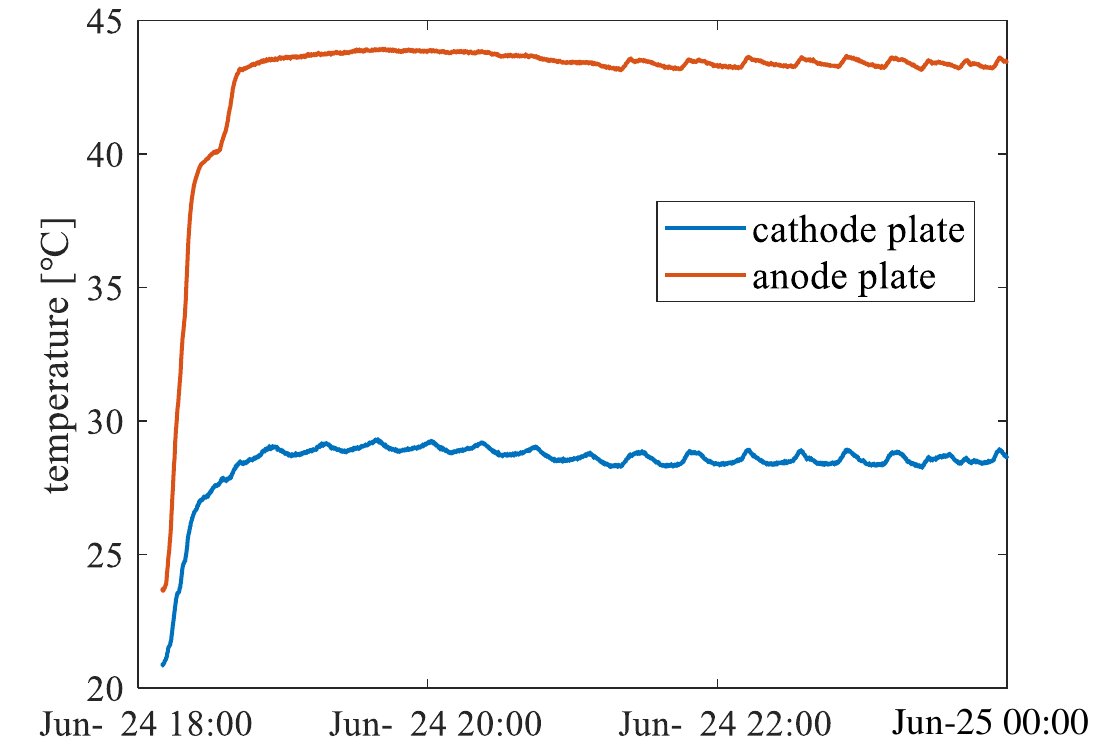}
\end{center}
\caption{The temperature change of the cathode and anode plates within 18 hours of continuous high power conditioning. The input power is 75 kW.}
\label{fig:temperature}
\end{figure}

Under 82~kW CW RF power, with a total water flow rate of approximately 248~L/min, the gun's resonant frequency decreased by about 100~kHz, exceeding the simulated prediction. Under normal water-cooling flow conditions, each tuner requires approximately 12~kN of force to adjust the resonant frequency to 216.667~MHz. Although each tuner can provide more than 15~kN of force, operating under such high force is detrimental to the gun's operating condition. Therefore, we reduced the water flow rate in the four rings of the anode plate from 10~L/min to approximately 5.5~L/min (there are five water channels in total, with the two outermost ones forming a semicircle). This caused the anode plate temperature to rise, and high-power testing revealed that the gun's resonant frequency increased. Figure~\ref{fig:temperature} shows the temperature variation of the cathode and anode plates during high-power conditioning. The cathode plate temperature stabilized at approximately 28~$^\circ$C, and the anode plate temperature stabilized at approximately 43~$^\circ$C. The aforementioned vacuum level of $2.2\times10^{-8}$~Pa under RF power was achieved at this elevated anode plate temperature. Therefore, if the resonant frequency of the next electron gun is better tuned during the assembly phase, a better vacuum level is expected.

\begin{figure}[htbp]
\begin{center}
\includegraphics[width=9cm]{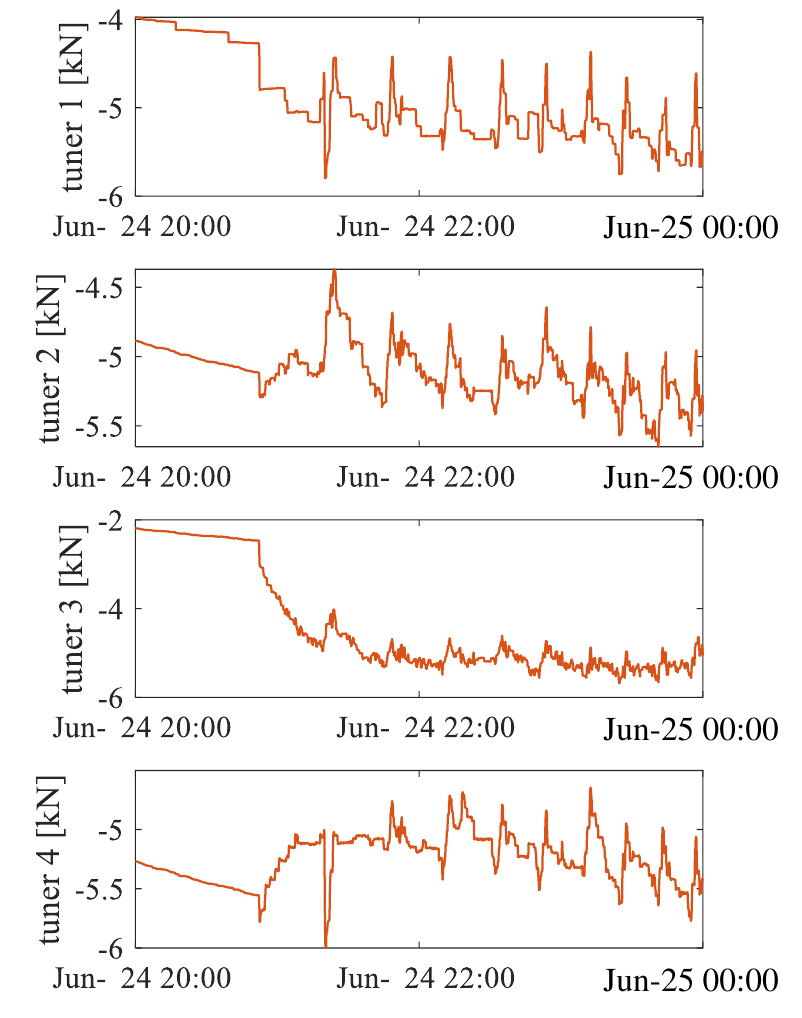}
\end{center}
\caption{Time-dependent force of the four tuners under 82 kW CW RF power.}
\label{fig:tuner}
\end{figure}

After adjusting the water flow through the anode plate, the evolution of the forces of the four tuners over time is shown in Fig.~\ref{fig:tuner}. The forces of the four tuners are controlled via a PLC cabinet. Our control algorithm incorporates a force-balancing mechanism. As can be seen, although the forces of the four tuners were initially inconsistent, they tended to become consistent after a period of operation. Eventually, the forces of the four tuners stabilized at approximately 5.5~kN.

\begin{figure*}[htbp]
\begin{center}
\includegraphics[width=14cm]{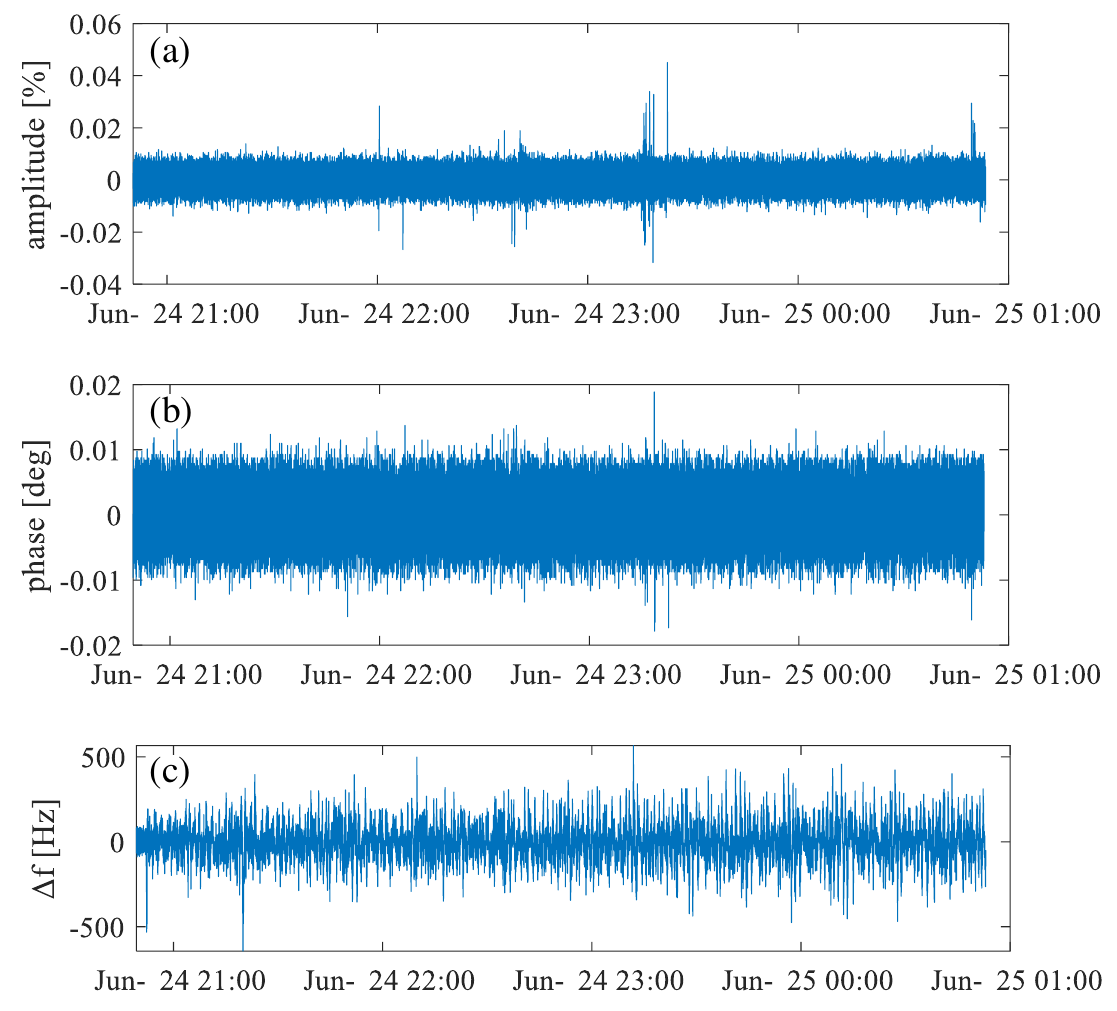}
\end{center}
\caption{ (a) The change of the amplitude of the probe signal. (b) The change of the phase of the probe signal. (c) The change of the gun resonant frequency.  The input power is 82kW.}
\label{fig:frequency_stability}
\end{figure*}

 The amplitude and phase stability of the electron gun during operation were characterized, as shown in Fig.~\ref{fig:frequency_stability}(a) and (b). The amplitude and phase of the microwaves coupled from the gun pickup were continuously monitored and controlled in a closed loop using a low-level RF system. The rms jitter of the amplitude was $3.32\times10^{-5}$, and the rms jitter of the phase was 0.0033$^\circ$. The gun resonance frequency was also continuously monitored via the low-level RF system, as shown in Fig.~\ref{fig:frequency_stability}(c). During tuner operation, the rms jitter of the resonance frequency was 104.8~Hz. Resonance frequency stability and amplitude–phase stability are closely related to fluctuations in the cooling water temperature. As shown in Fig.~\ref{fig:temperature}, the temperature fluctuations of the cathode and anode plates under steady-state conditions are $\pm 0.4~^\circ$C, which is consistent with the water temperature fluctuations in our cooling system. The $\rm S^3$FEL facility in Shenzhen will feature a more advanced cooling water system, with water temperature fluctuations expected to be within $\pm 0.1~^\circ$C. Therefore, once the electron gun is installed at the $\rm S^3$FEL, we anticipate smaller fluctuations in the resonance frequency and improved amplitude and phase stability.

\begin{figure}[htbp]
\begin{center}
\includegraphics[width=9cm]{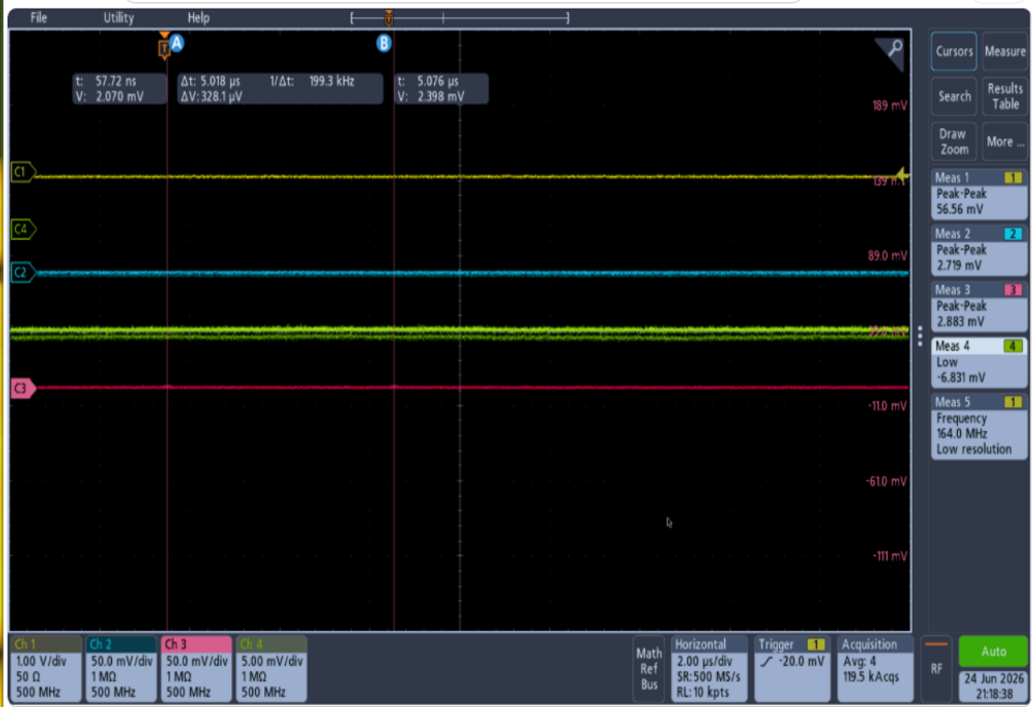}
\end{center}
\caption{Oscilloscope signals with RF power of 82 kW. The blue and red lines are the current signals detected by the micro-current monitors mounted on the two couplers, respectively. The green line denotes the dark current signal. The dark current was measured to be 6.8 nA.}
\label{fig:darkcurrent}
\end{figure}

 As shown in Fig.~\ref{fig:high_power_test_image}, a Faraday cup was installed at the electron gun outlet to measure the dark current. A SAES NexTorr NEG-ion Z200 pump was installed near the Faraday cup to maintain a good vacuum inside it. The current signal from the Faraday cup was transmitted to an oscilloscope via a coaxial cable. The oscilloscope measurement interface is shown in Fig.~\ref{fig:darkcurrent}. The green trace represents the dark current signal. The dark current was measured to be 6.8~nA under 82~kW of RF power. In addition, the blue and red traces correspond to the current signals detected by the microcurrent monitors mounted on the two couplers, respectively. When operating in CW mode at 82~kW input power, the current signals from both couplers disappeared, indicating that no multipacting occurred inside the coupler waveguide at this time.

Based on Table~\ref{tab:table2}, under the condition of 0.5~mm over-insertion of the plug, with an unloaded quality factor $Q_0' = 34{,}533$ and an input power $P' = 76.2$~kW, the voltage $V' = 854$~kV and the cathode electric field gradient $E' = 29.86$~MV/m. In the experiment, the unloaded quality factor is $Q_0^{\text{ex}} = 32{,}255$, and the input power is $P^{\text{ex}} = 82$~kW. Accordingly, the voltage should be $V^{\text{ex}} = V' \sqrt{\dfrac{P^{\text{ex}} Q_0^{\text{ex}}}{P' Q_0'}} = 856$~kV, and the cathode electric field gradient should be $E^{\text{ex}} = E' \sqrt{\dfrac{P^{\text{ex}} Q_0^{\text{ex}}}{P' Q_0'}} = 29.9$~MV/m.

\section{Conclusion}

A VHF photocathode electron gun with a resonant frequency of 216.667~MHz was developed for the $\rm S^3$FEL facility. By optimizing the cavity profile, multipacting behavior, and cooling design, the new gun achieves a higher voltage and cathode gradient than the SHINE/DALS gun at comparable input power. During high-power conditioning, the gun was successfully operated with 82~kW of CW RF power, reaching a cathode gradient of 29.9~MV/m and a voltage of 856~kV, again exceeding the previous world record for room-temperature VHF guns. The dark current measured at the gun exit was only 6.8~nA. The stability of the amplitude, phase, and resonance frequency was also characterized. The rms amplitude jitter was $3.32\times10^{-5}$, and the rms phase jitter was 0.0033$^\circ$. The rms resonance frequency jitter was 104.8~Hz. The vacuum inside the gun was maintained at $4\times10^{-9}$~Pa without RF power and at $2.2\times10^{-8}$~Pa with an RF power of 82~kW. These results demonstrate that the optimized VHF gun is ready for deployment as the electron source for $\rm S^3$FEL and provides a solid foundation for future high-repetition-rate FEL and UED applications.

\section*{Acknowledgments}
This work is supported by the National Natural Science Foundation of China (NSFC) under Grant No. 12275149.

\bibliography{mybibfile.bib}

\end{document}